# Temperature Dependence of Wavelength Selectable Zero-Phonon Emission from Single Defects in Hexagonal Boron Nitride

Nicholas R. Jungwirth,[1] Brian Calderon,[1] Yanxin Ji,[1] Michael G. Spencer,[1] Michael E. Flatté,[2] and Gregory D. Fuchs[1]

[1]Cornell University, Ithaca, New York 14853, USA

[2]University of Iowa, Iowa City, Iowa 52242, USA

**Abstract:**

We investigate the distribution and temperature-dependent optical properties of sharp, zero-phonon emission from defect-based single photon sources in multilayer hexagonal boron nitride (h-BN) flakes. We observe sharp emission lines from optically active defects distributed across an energy range that exceeds 500 meV. Spectrally-resolved photon-correlation measurements verify single photon emission, even when multiple emission lines are simultaneously excited within the same h-BN flake. We also present a detailed study of the temperature-dependent linewidth, spectral energy shift, and intensity for two different zero-phonon lines centered at 575 nm and 682 nm, which reveals a nearly identical temperature dependence despite a large difference in transition energy. Our temperature-dependent results are best described by a lattice vibration model that considers piezoelectric coupling to in-plane phonons. Finally, polarization spectroscopy measurements suggest that whereas the 575 nm emission line is directly excited by 532 nm excitation, the 682 nm line is excited indirectly.

Two-dimensional materials and associated layered solids including graphene, hexagonal boron nitride (h-BN), and transition metal dichalcogenides (TMDs) possess attractive mechanical,[1,2] electrical,[3,4] thermal,[5,6] chemical,[7,8] and optical[9,10] properties. Unlike graphene and TMDs, h-BN is a wide bandgap (~6 eV) electrical insulator, making it a key component in many van der Waals heterostructures.[11–13] This feature also makes h-BN an ideal host for optically active defect centers.[14,15]

Isolated color centers in wide bandgap semiconductors are single photon sources with potential applications in quantum optics, precision sensing, and quantum information processing technologies.[14,16–19] Recently, ultrabright and polarized single photon emission from isolated defects in monolayer and multilayer h-BN has been reported.[20–22] These observations add h-BN to the growing collection of wide bandgap materials (Diamond,[23–31] SiC,[32–38] and ZnO[15,39–43]) that host defect-based room temperature single photon sources. Whereas isolated defects in monolayer h-BN show broad spectral emission and unreliable photostability in our measurements (see Supporting Information), single defects in multilayer h-BN appear to be absolutely photostable and exhibit sharp zero-phonon lines with a small Huang-Rhys factor.[44–46] These properties potentially make multilayer h-BN defects attractive sources of indistinguishable single photons. At present, defect centers in both monolayer and multilayer h-BN remain poorly understood, which motivates investigation of the properties of defects within each material.

In this work we study the temperature dependence of spectrally narrow zero-phonon lines (ZPLs) from point defects in multilayer h-BN. First, we characterize the distribution and intensity of sharp spectral emission in the range ~570-740 nm at cryogenic temperatures. This survey reveals a forest of sharp spectral features across a wide range of energy. Spectrally-resolved photon-correlation measurements enable us to unambiguously identify features that correspond to the ZPLs of individual defects. For two of the identified ZPLs (575 nm and 682 nm) we investigate the temperature-dependent linewidth, line shift, excited state lifetime, and intensity. Despite an energy difference of over 300 meV, both ZPLs share similar temperature-dependent line broadening and shifting. We propose a phonon-mediated mechanism to explain these observations. Finally, we present polarization spectroscopy results that suggest while the 575 nm line is excited directly through defect absorption of the 532 nm exciting light, the 682 nm line is excited indirectly though cross relaxation.

**Results and Discussion:**

A house-built confocal microscope was used to selectively excite deep defect levels in h-BN flakes (See Experimental for sample information and

Supporting Information for microscope details). Flakes possessing optically active defects produced bright fluorescence (>400 kPhotons/s) when excited with 532 nm light. Figure 1a displays representative fluorescence spectra of two distinct h-BN flakes at 4.5 K. Each spectrum possesses several sharp emission lines, some of which may correspond to the ZPL of an individual defect. In Figure 1b we display the two-photon correlation function, $g^{(2)}(\tau)$, measured on the h-BN flake that produced the red spectrum in Figure 1a. For this measurement collection was limited to the shaded spectral region of 675-700 nm with optical filters. Because the antibunching dip at $\tau = 0$ extends to ~0.15, the sharp spectral feature at 682 nm marked with an arrow can be identified as single photon emission. We note that quantum emission into a narrow line likely corresponds to both zero-phonon and low-energy phonon-mediated emission from an individual defect but, in principle, could correspond to a sharp optical-phonon sideband of a higher energy ZPL. However, the temperature dependence of the feature's intensity and linewidth provide strong support that at low temperature the feature largely originates from zero-phonon processes. Thus we take sharp emission lines with $g^{(2)}(0) < 0.5$ to be ZPLs of individual defects.

    The energy of a ZPL is determined primarily by the Hamiltonian describing the orbital component of the defect wavefunction. However, the orbital excited-to-ground-state splitting may be shifted by defect-to-defect variations in the local environment such as strain or trapped charges. To investigate this variation in multilayer h-BN, we collected emission spectra similar to those shown in Figure

1a from 90 distinct h-BN flakes at 4.5 K. We restricted our attention to spectral features with a full width at half maximum (FWHM) of less than 1 nm. We recorded the peak position ($E_0$) and relative intensity ($I_R$) of each feature, where $I$ is the ratio of the spectral intensity at $E_0$ to that of the background (see Supporting Information). Figure 2a is a scatter plot relating $E_0$ to $I_R$ for the 340 lines we identified. We only include emission lines with $I_R > 2$ because only these lines may in principle produce $g^{(2)}(0) < 0.5$. Emission lines in h-BN can be bright ($I_R \approx 100$), even at photon energies far removed from the excitation source energy.

Figure 2b is a histogram of the positions of the lines in Figure 2a. Several regions of clustering are evident and each cluster may result from the ZPL of a unique defect species. For example, a previous report[20] identified a ZPL with energy ~1.99 eV that was tentatively attributed to an anti-site nitrogen vacancy $N_BV_N$. Each clustered region may be inhomogeneously broadened by variations in the local strain that serve to shift individual ZPL positions. Though we did not measure $g^{(2)}(\tau)$ for every narrow emission line, the green hexagons above the x-axis correspond to ZPLs that we have spectrally isolated with optical filters and measured $g^{(2)}(0) < 0.5$. Therefore, we conclude from Figure 2 that h-BN flakes host multiple species of bright, photostable defects with ZPL energies that are selectable over a broad range.

The distribution of ZPL positions in Figure 2 suggest that the defects couple to local lattice strain. To better understand defect-lattice interactions, and specifically dynamical (phonon-mediated) interactions, we investigated the temperature dependence of two ZPLs located at 575 nm (2.16 eV) and 682 nm (1.82 eV). For each temperature investigated, we fit the emission spectrum to a linear combination of a background and a Lorentzian lineshape,

$$L(E) \propto \frac{1}{(E-E_{ZPL})^2 + (\Gamma/2)^2}, \tag{1}$$

where $E_{ZPL}$ is the ZPL energy, $\Gamma$ is the FWHM, and $L(E)dE$ is proportional to the number of photons emitted in the energy range $(E, E + dE)$. Our results for $\Delta E_{ZPL}(T) = E_{ZPL}(T) - E_{ZPL}(4.5\ K)$ and $\Gamma(T)$ are plotted in Figure 3a and b, respectively. We note from Figure 3a that both ZPLs red shift with increasing temperature in a similar manner, despite having transition energies that differ by ~340 meV. The solid line is a guide to the eye with a $T^3$ trend. In Figure 3b the FWHM broadens with temperature and is spectrometer-resolution limited below ~15 K. The inset, which is the same data plotted on a log scale, indicates a nearly exponential temperature dependence. The trends in Figure 3ab differ from those reported in other defect system[25,28,38] and from theoretical expectations for defects in a 3D solid.[45] To better understand our observations we consider both lifetime- and phonon-mediated broadening mechanisms below.

First we compare the observed linewidth to the minimal, or natural, linewidth for a transition, which in the absence of phonons is given by the expression $\Gamma_{min} = \hbar/\tau$, where $\hbar$ is the reduced Planck's constant. This is the Lorentzian linewidth that is limited by the Fourier transform of spontaneous emission, with a characteristic excited-state lifetime $\tau$. To determine whether the broadening we observe is related to a change in the natural linewidth, we also measured the temperature-dependent excited-state lifetime for each ZPL transition using pulsed excitation. This measurement, shown in Figure 3c, indicates that the lifetime is independent of temperature, and that the natural linewidths are 0.352 $\mu eV$ and 0.148 $\mu eV$ for the 575 nm line (ZPL1) and the 682 nm line (ZPL2), respectively. These are much narrower than the observed linewidths, which we note are not spectrometer-limited for the majority of temperatures. Therefore, we conclude that the linewidth broadening is not caused by a change in the natural linewidth, and that a different temperature-dependent process dominates the linewidth between 4.5 and 360 K.

Here we propose a phonon-mediated mechanism to explain the exponential linewidth broadening evident in Figure 3b. In a model with a single phonon frequency $\omega_0$, interactions with phonons produce one-phonon and multiphonon sidebands that emerge as distinct satellite peaks.[44] In our model of h-BN defect emission, however, low-energy acoustic phonons, which are ungapped,[45] produce an effective increased width to the zero-phonon line. The one-acoustic-phonon sideband is peaked at vanishing phonon energy due to the Bose factor governing the phonon occupation and has a width that increases approximately linearly with temperature. Photons emitted into the one-acoustic-phonon sideband are thus very close in energy to the zero-phonon transition energy and are experimentally indistinguishable from the "true" zero-phonon line in our optical spectrometer. At the same time, the spectral weight of the zero-phonon line diminishes with respect to the total emission according to the temperature-dependent factor

$$W = \exp[-S \coth(\hbar\omega_0/2kT)], \tag{2}$$

where $k$ is Boltzmann's constant and $S$ is the Huang-Rhys factor.[45] Thus, the ratio of the amplitude of the one-phonon contribution to the zero-phonon contribution of the narrow line will increase nearly exponentially with temperature, yielding an effective linewidth for the observed line that likewise increases exponentially with temperature. The solid lines in Figure 3b are best-fits to the data using this model with two free parameters. The best-fit assumes piezoelectric coupling between the defect and in-plane acoustic phonons (see Supporting Information). Incidentally, the defect predominantly couples to phonons in its particular two-dimensional h-BN sheet. The Huang-Rhys factors from the best-fits are $S = 2.1 \pm 0.7$ and $S = 3.3 \pm 2.3$ for ZPL1 and ZPL2, respectively. We note that the ZPL factor $W$ in Equation 2 relates to the relative spectral weight of the ZPL.[46] Our Huang-Rhys factors correspond to 0 K ZPL factors of $W = 0.12$ and $W = 0.037$ for ZPL1 and ZPL2, respectively, consistent with the sharp, bright spectral lines experimentally observed. These values are much less than those reported previously for isolated h-BN defects[20] because our model explicitly accounts for low energy acoustic phonon processes that cannot be experimentally distinguished from zero-phonon processes by a spectrometer.

Because ZPL2 has a larger Huang-Rhys factor, we expect the amplitude of ZPL2 to decrease more rapidly with temperature than ZPL1. This expectation is verified in Figure 3d, which displays the temperature dependence of the relative intensity of each ZPL (see Supporting Information). While the relative intensity of each line decays nearly exponentially with temperature, ZPL2 decays more rapidly than does ZPL1. This observation explains why ZPL1 was studied over a broader temperature range than ZPL2.

To further examine the excitation mechanism of each ZPL, we performed polarization spectroscopy[42] on ZPL1 (Figure 4a) and ZPL2 (Figure 4b). ZPL1 possesses large absorption and emission polarization visibility with the maxima of each profile being aligned. This behavior is consistent with a single absorption and emission dipole directly coupling the defect's ground and excited state wavefunctions. Moreover, group theoretical considerations suggest that the ground and excited state wavefunctions of ZPL1 are likely orbital singlets.[42,47] Conversely, ZPL2 shows lower polarization visibility with a ~60° misalignment between the maxima of absorption and emission. These observations suggest that the electronic states coupled by the incident laser field differ from those that produce ZPL2, perhaps because ZPL2 results from cross relaxation between defects.

**Conclusion:**

We investigated the distribution and temperature dependence of sharp emission lines in h-BN to better understand the properties of native point defects. Our results, which survey a range of ~500 meV, reveal a dense forest of narrow emission lines. Though dense and broad, the distribution of spectral features shows regions of clustering, where each cluster may correspond to the ZPL of a unique defect species. By combining optical filters with time correlated single photon counting, we identify six narrow lines that correspond to single photon, zero-phonon emission from individual defect states.

For two of the identified ZPLs, we investigate the temperature dependent linewidth, energy shift, and intensity. Each ZPL exhibits similar linewidth broadening and red shifting as temperature is increased. In contrast to other defect systems in diamond[25,28] and SiC,[38] the measured linewidth increases exponentially with temperature. We explain this result by invoking the nature of the in-plane one-acoustic-phonon sideband, and speculate that this feature is more pronounced for the h-BN lattice due to its van der Waals nature (softer modes). Additionally, we find that the relative intensity of each ZPL decreases exponentially with temperature. Finally we present evidence that h-BN defects may either be excited by direct phonon-mediated absorption, or indirectly by cross-relaxation.

**Experimental:**

The investigated h-BN flakes are commercially available from Graphene Supermarket (See Supporting Information for sample characterization). As received the flakes are suspended in an ethanol/water solution. After drop casting $25\ \mu L$ of solution onto a thermally oxidized Si substrate, we anneal samples at $850°\ C$ for 30 minutes under continuous nitrogen flow. The ramp rate was $5°\ C/min$ and the samples were allowed to cool overnight after the annealing treatment. All measurements were performed on samples loaded in a Janis ST-500 cryostat using a house-built confocal microscope (See Supporting Information). When collecting temperature dependent data, a minimum of one minute was allotted for each degree Kelvin temperature change to ensure that the sample had reached thermal equilibrium.

**Acknowledgements:**

This work was supported by the National Science Foundation (DMR-1254530). We acknowledge use of the Cornell NanoScale Facility, a member of the National Nanotechnology Coordinated Infrastructure (NNCI), which is supported by the National Science Foundation (Grant ECCS-15420819). Additionally, we acknowledge the Cornell Center for Materials Research Shared Facilities, which are supported through the NSF MRSEC program (DMR-1120296). M. E. F. acknowledges support from an AFOSR MURI.

**References:**

(1) Lee, G.-H.; Cooper, R. C.; An, S. J.; Lee, S.; van der Zande, A.; Petrone, N.; Hammerberg, A. G.; Lee, C.; Crawford, B.; Oliver, W.; *et al.* High-Strength Chemical-Vapor–deposited Graphene and Grain Boundaries. *Science* **2013**, *340*, 1073–1076.
(2) Shekhawat, A.; Ritchie, R. O. Toughness and Strength of Nanocrystalline Graphene. *Nat. Commun.* **2016**, *7*, 10546.
(3) Bolotin, K. I.; Sikes, K. J.; Jiang, Z.; Klima, M.; Fudenberg, G.; Hone, J.; Kim, P.; Stormer, H. L. Ultrahigh Electron Mobility in Suspended Graphene. *Solid State Commun.* **2008**, *146*, 351–355.
(4) Baringhaus, J.; Ruan, M.; Edler, F.; Tejeda, A.; Sicot, M.; Taleb-Ibrahimi, A.; Li, A.-P.; Jiang, Z.; Conrad, E. H.; Berger, C.; *et al.* Exceptional Ballistic Transport in Epitaxial Graphene Nanoribbons. *Nature* **2014**, *506*, 349–354.
(5) Chen, C.-C.; Li, Z.; Shi, L.; Cronin, S. B. Thermal Interface Conductance across a Graphene/hexagonal Boron Nitride Heterojunction. *Appl. Phys. Lett.* **2014**, *104*, 081908.
(6) Sakhavand, N.; Shahsavari, R. Dimensional Crossover of Thermal Transport in Hybrid Boron Nitride Nanostructures. *ACS Appl. Mater. Interfaces* **2015**, *7*, 18312–18319.

(7) Lei, W.; Portehault, D.; Liu, D.; Qin, S.; Chen, Y. Porous Boron Nitride Nanosheets for Effective Water Cleaning. *Nat. Commun.* **2013**, *4*, 1777.
(8) Lian, G.; Zhang, X.; Si, H.; Wang, J.; Cui, D.; Wang, Q. Boron Nitride Ultrathin Fibrous Nanonets: One-Step Synthesis and Applications for Ultrafast Adsorption for Water Treatment and Selective Filtration of Nanoparticles. *ACS Appl. Mater. Interfaces* **2013**, *5*, 12773–12778.
(9) Kim, C.-J.; Sánchez-Castillo, A.; Ziegler, Z.; Ogawa, Y.; Noguez, C.; Park, J. Chiral Atomically Thin Films. *Nat. Nanotechnol* **2016**.
(10) Kumar, A.; Nemilentsau, A.; Fung, K. H.; Hanson, G.; Fang, N. X.; Low, T. Chiral Plasmon in Gapped Dirac Systems. *Phys. Rev. B* **2016**, *93*.
(11) Geim, A. K.; Grigorieva, I. V. Van Der Waals Heterostructures. *Nature* **2013**, *499*, 419–425.
(12) Wang, X.; Xia, F. Van Der Waals Heterostructures: Stacked 2D Materials Shed Light. *Nat. Mater.* **2015**, *14*, 264–265.
(13) Cui, X.; Lee, G.-H.; Kim, Y. D.; Arefe, G.; Huang, P. Y.; Lee, C.-H.; Chenet, D. A.; Zhang, X.; Wang, L.; Ye, F.; *et al.* Multi-Terminal Transport Measurements of MoS2 Using a van Der Waals Heterostructure Device Platform. *Nat. Nanotechnol.* **2015**, *10*, 534–540.
(14) Weber, J. R.; Koehl, W. F.; Varley, J. B.; Janotti, A.; Buckley, B. B.; Van de Walle, C. G.; Awschalom, D. D. Quantum Computing with Defects. *Proc. Natl. Acad. Sci.* **2010**, *107*, 8513–8518.
(15) Jungwirth, N. R.; Pai, Y. Y.; Chang, H. S.; MacQuarrie, E. R.; Nguyen, K. X.; Fuchs, G. D. A Single-Molecule Approach to ZnO Defect Studies: Single Photons and Single Defects. *J. Appl. Phys.* **2014**, *116*, 043509.
(16) Kucsko, G.; Maurer, P. C.; Yao, N. Y.; Kubo, M.; Noh, H. J.; Lo, P. K.; Park, H.; Lukin, M. D. Nanometre-Scale Thermometry in a Living Cell. *Nature* **2013**, *500*, 54–58.
(17) Maze, J. R.; Stanwix, P. L.; Hodges, J. S.; Hong, S.; Taylor, J. M.; Cappellaro, P.; Jiang, L.; Dutt, M. V. G.; Togan, E.; Zibrov, A. S.; *et al.* Nanoscale Magnetic Sensing with an Individual Electronic Spin in Diamond. *Nature* **2008**, *455*, 644–647.
(18) Toyli, D. M.; de las Casas, C. F.; Christle, D. J.; Dobrovitski, V. V.; Awschalom, D. D. Fluorescence Thermometry Enhanced by the Quantum Coherence of Single Spins in Diamond. *Proc. Natl. Acad. Sci.* **2013**, *110*, 8417–8421.
(19) Chu, Y.; Lukin, M. D. Quantum Optics with Nitrogen-Vacancy Centers in Diamond. *ArXiv Prepr. ArXiv150405990* **2015**.
(20) Tran, T. T.; Bray, K.; Ford, M. J.; Toth, M.; Aharonovich, I. Quantum Emission from Hexagonal Boron Nitride Monolayers. *Nat. Nanotechnol.* **2015**, *11*, 37–41.
(21) Tran, T. T.; Zachreson, C.; Berhane, A. M.; Bray, K.; Sandstrom, R. G.; Li, L. H.; Taniguchi, T.; Watanabe, K.; Aharonovich, I.; Toth, M. Quantum Emission from Defects in Single-Crystalline Hexagonal Boron Nitride. *Phys. Rev. Appl.* **2016**, *5*.
(22) Tran, T. T.; ElBadawi, C.; Totonjian, D.; Lobo, C. J.; Grosso, G.; Moon, H.; Englund, D. R.; Ford, M. J.; Aharonovich, I.; Toth, M. Robust Multicolor Single Photon Emission from Point Defects in Hexagonal Boron Nitride. *ArXiv Prepr. ArXiv160309608* **2016**.

(23) Kurtsiefer, C.; Mayer, S.; Zarda, P.; Weinfurter, H. Stable Solid-State Source of Single Photons. *Phys. Rev. Lett.* **2000**, *85*, 290–293.
(24) Brouri, R.; Beveratos, A.; Poizat, J.-P.; Grangier, P. Photon Antibunching in the Fluorescence of Individual Color Centers in Diamond. *Opt. Lett.* **2000**, *25*, 1294–1296.
(25) Fu, K.-M. C.; Santori, C.; Barclay, P. E.; Rogers, L. J.; Manson, N. B.; Beausoleil, R. G. Observation of the Dynamic Jahn-Teller Effect in the Excited States of Nitrogen-Vacancy Centers in Diamond. *Phys. Rev. Lett.* **2009**, *103*.
(26) Simpson, D. A.; Ampem-Lassen, E.; Gibson, B. C.; Trpkovski, S.; Hossain, F. M.; Huntington, S. T.; Greentree, A. D.; Hollenberg, L. C. L.; Prawer, S. A Highly Efficient Two Level Diamond Based Single Photon Source. *Appl. Phys. Lett.* **2009**, *94*, 203107.
(27) Zhao, H.-Q.; Fujiwara, M.; Okano, M.; Takeuchi, S. Observation of 12-GHz Linewidth of Zero-Phonon-Line in Photoluminescence Spectra of Nitrogen Vacancy Centers in Nanodiamonds Using a Fabry-Perot Interferometer. *Opt. Express* **2013**, *21*, 29679.
(28) Neu, E.; Hepp, C.; Hauschild, M.; Gsell, S.; Fischer, M.; Sternschulte, H.; Steinmüller-Nethl, D.; Schreck, M.; Becher, C. Low-Temperature Investigations of Single Silicon Vacancy Colour Centres in Diamond. *New J. Phys.* **2013**, *15*, 043005.
(29) Gao, W. B.; Imamoglu, A.; Bernien, H.; Hanson, R. Coherent Manipulation, Measurement and Entanglement of Individual Solid-State Spins Using Optical Fields. *Nat. Photonics* **2015**, *9*, 363–373.
(30) Leifgen, M.; Schröder, T.; Gädeke, F.; Riemann, R.; Métillon, V.; Neu, E.; Hepp, C.; Arend, C.; Becher, C.; Lauritsen, K.; *et al.* Evaluation of Nitrogen- and Silicon-Vacancy Defect Centres as Single Photon Sources in Quantum Key Distribution. *New J. Phys.* **2014**, *16*, 023021.
(31) Hensen, B.; Bernien, H.; Dréau, A. E.; Reiserer, A.; Kalb, N.; Blok, M. S.; Ruitenberg, J.; Vermeulen, R. F. L.; Schouten, R. N.; Abellán, C.; *et al.* Experimental Loophole-Free Violation of a Bell Inequality Using Entangled Electron Spins Separated by 1.3 Km. *ArXiv Prepr. ArXiv150805949* **2015**.
(32) Weber, J. R.; Koehl, W. F.; Varley, J. B.; Janotti, A.; Buckley, B. B.; Van de Walle, C. G.; Awschalom, D. D. Defects in SiC for Quantum Computing. *J. Appl. Phys.* **2011**, *109*, 102417.
(33) Koehl, W. F.; Buckley, B. B.; Heremans, F. J.; Calusine, G.; Awschalom, D. D. Room Temperature Coherent Control of Defect Spin Qubits in Silicon Carbide. *Nature* **2011**, *479*, 84–87.
(34) Castelletto, S.; Johnson, B. C.; Ivády, V.; Stavrias, N.; Umeda, T.; Gali, A.; Ohshima, T. A Silicon Carbide Room-Temperature Single-Photon Source. *Nat. Mater.* **2013**.
(35) Widmann, M.; Lee, S.-Y.; Rendler, T.; Son, N. T.; Fedder, H.; Paik, S.; Yang, L.-P.; Zhao, N.; Yang, S.; Booker, I.; *et al.* Coherent Control of Single Spins in Silicon Carbide at Room Temperature. *Nat. Mater.* **2014**, *14*, 164–168.
(36) Christle, D. J.; Falk, A. L.; Andrich, P.; Klimov, P. V.; Hassan, J. U.; Son, N. T.; Janzén, E.; Ohshima, T.; Awschalom, D. D. Isolated Electron Spins in Silicon Carbide with Millisecond Coherence Times. *Nat. Mater.* **2014**, *14*, 160–163.

(37) Fuchs, F.; Stender, B.; Trupke, M.; Simin, D.; Pflaum, J.; Dyakonov, V.; Astakhov, G. V. Engineering near-Infrared Single-Photon Emitters with Optically Active Spins in Ultrapure Silicon Carbide. *Nat. Commun.* **2015**, *6*, 7578.

(38) Lienhard, B.; Schröder, T.; Mouradian, S.; Dolde, F.; Tran, T. T.; Aharonovich, I.; Englund, D. R. Bright and Stable Visible-Spectrum Single Photon Emitter in Silicon Carbide. *ArXiv Prepr. ArXiv160305759* **2016**.

(39) Morfa, A. J.; Gibson, B. C.; Karg, M.; Karle, T. J.; Greentree, A. D.; Mulvaney, P.; Tomljenovic-Hanic, S. Single-Photon Emission and Quantum Characterization of Zinc Oxide Defects. *Nano Lett.* **2012**, *12*, 949–954.

(40) Choi, S.; Johnson, B. C.; Castelletto, S.; Ton-That, C.; Phillips, M. R.; Aharonovich, I. Single Photon Emission from ZnO Nanoparticles. *Appl. Phys. Lett.* **2014**, *104*, 261101.

(41) Neitzke, O.; Morfa, A.; Wolters, J.; Schell, A. W.; Kewes, G.; Benson, O. Investigation of Line Width Narrowing and Spectral Jumps of Single Stable Defect Centers in ZnO at Cryogenic Temperature. *Nano Lett.* **2015**, *15*, 3024–3029.

(42) Jungwirth, N. R.; Chang, H.-S.; Jiang, M.; Fuchs, G. D. Polarization Spectroscopy of Defect-Based Single Photon Sources in ZnO. *ACS Nano* **2016**, *10*, 1210–1215.

(43) Ghosh, S.; Ghosh, M.; Seibt, M.; Mohan Rao, G. Detection of Quantum Well Induced Single Degenerate-Transition-Dipoles in ZnO Nanorods. *Nanoscale* **2016**, *8*, 2632–2638.

(44) Huang, K.; Rhys, A. Theory of Light Absorption and Non-Radiative Transitions in F-Centres. *Proc R. Soc* **1950**, *204*.

(45) Stoneham, A. M. *Theory of Defects in Solids: Electronic Structure of Defects in Insulators and Semiconductors*; OUP: Oxford, 2001.

(46) Alkauskas, A.; Buckley, B. B.; Awschalom, D. D.; Van de Walle, C. G. First-Principles Theory of the Luminescence Lineshape for the Triplet Transition in Diamond NV Centres. *New J. Phys.* **2014**, *16*, 073026.

(47) Dresselhaus, M. S.; Dresselhaus, G.; Jorio, A. *Group Theory: Application to the Physics of Condensed Matter*; SpringerLink: Springer e-Books; Springer Berlin Heidelberg, 2007.

**Figure 1:**

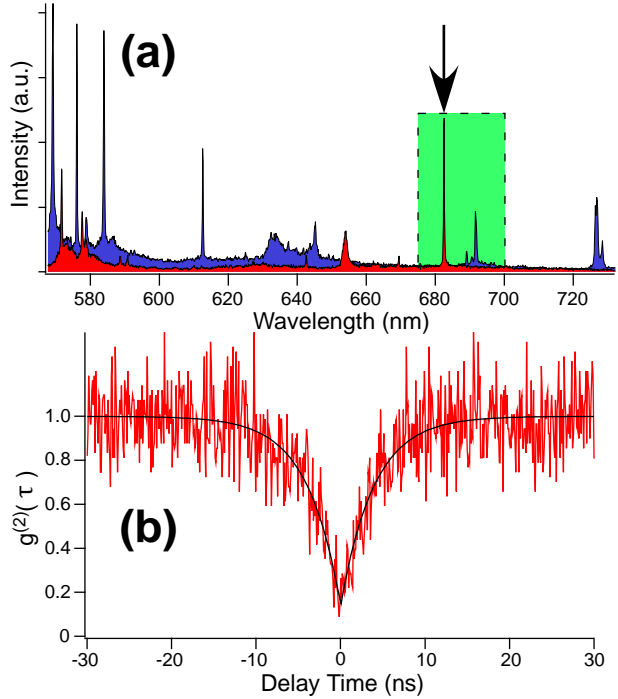

(a) Representative emission spectra for two distinct h-BN flakes revealing multiple sharp emission lines. (b) A $g^{(2)}(\tau)$ measurement of the red spectrum in (a), where collection was limited to the shaded region. The antibunching dip at $\tau = 0$ verifies single photon emission from the zero phonon line in (a) marked by the arrow.

**Figure 2:**

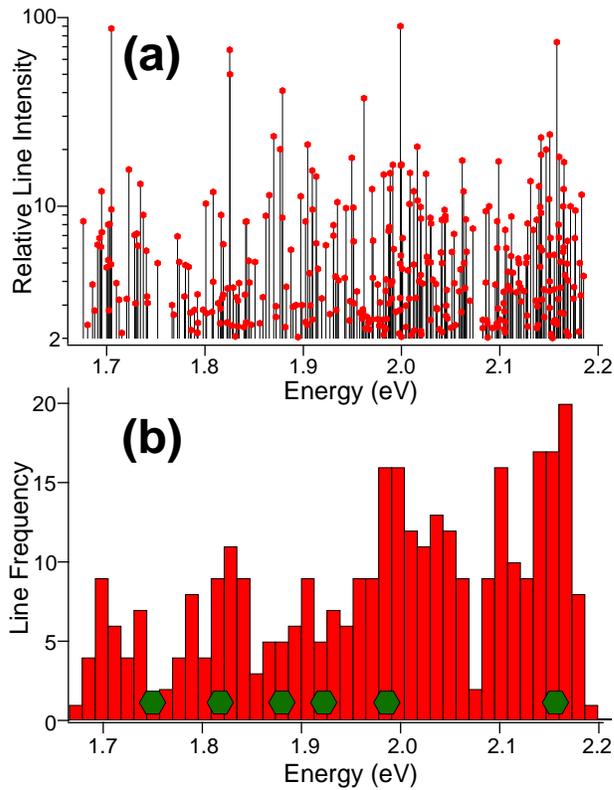

(a) Scatter plot relating energy and relative intensity of sharp emission lines from 90 distinct h-BN flakes. Individual emission lines may be bright and are densely distributed across a broad energy range. (b) The histogram of line positions from (a) reveals clustering, where each cluster may result from the ZPL of a unique defect. The green hexagons along the x-axis denote emission lines for which we measured $g^{(2)}(0) < 0.5$ using spectrally-resolved photon correlation measurements.

**Figure 3:**

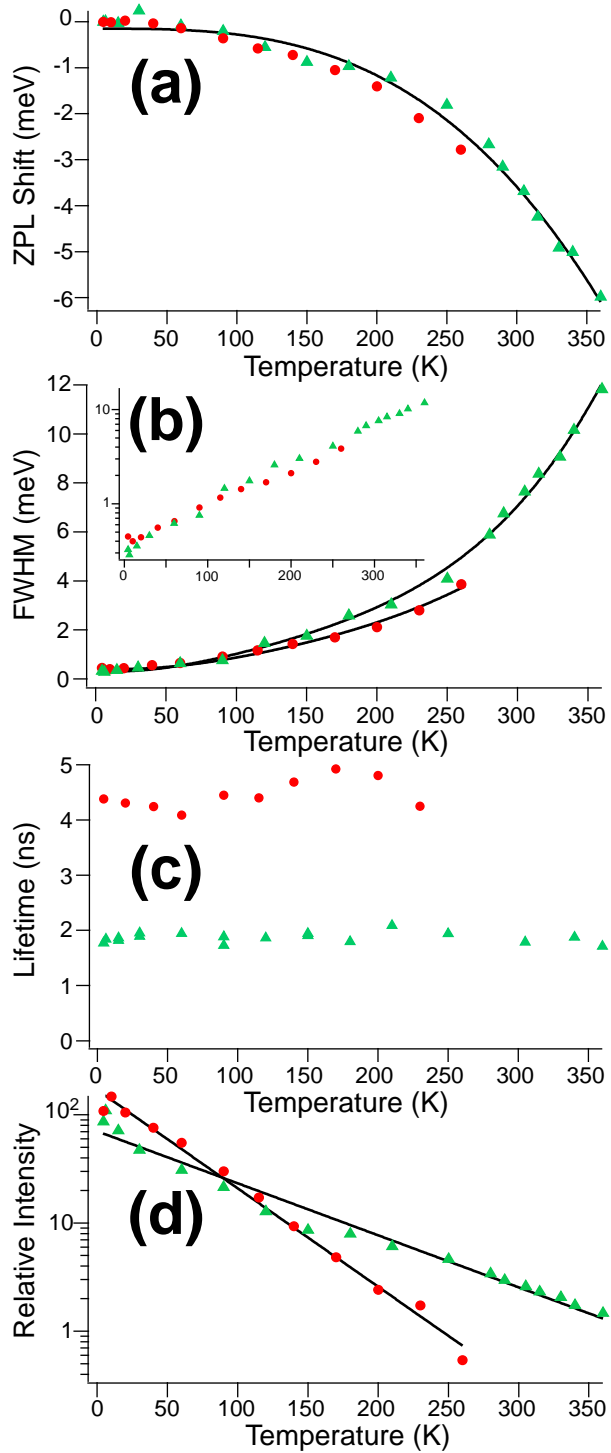

Temperature-dependent energy shift (a), linewidth (b), lifetime (c), and relative intensity (d) for two ZPLs centered at 575 nm (green triangles) and 682 nm (red circles). Each ZPL shows nearly identical behavior in (a)-(c), though the ZPL at 682 nm decreases in intensity more rapidly as temperature is increased.

**Figure 4:**

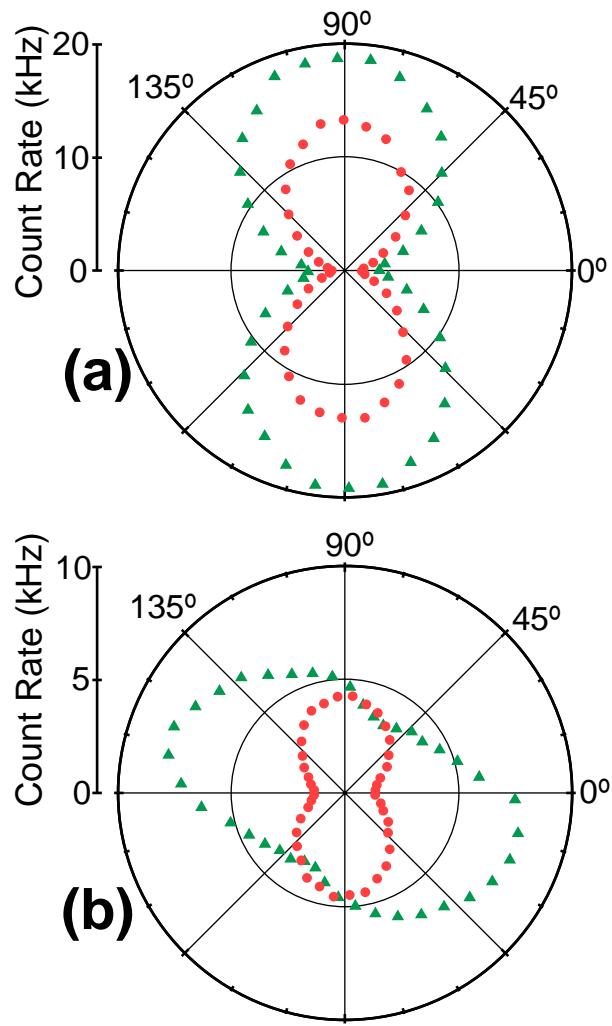

Polarization spectroscopy measurements of absorption (green triangles) and emission (red circles) for ZPLs at 575 nm (a) and 682 nm (b). The profiles in (a) are consistent with a single absorption and emission dipole aligned parallel to one another. The profiles in (b) show lower polarization visibility and the maxima of absorption and emission are misaligned by ~60°, suggesting the 682 nm line is excited indirectly.

# Supplemental Material: "Temperature Dependence of Wavelength Selectable Zero-Phonon Emission from Single Defects in Hexagonal Boron Nitride"

Nicholas R. Jungwirth,[1] Brian Calderon,[1] Yanxin Ji,[1] Michael G. Spencer,[1] Michael E. Flatté,[2] and Gregory D. Fuchs[1]

[1]Cornell University, Ithaca, New York 14853, USA

[2]University of Iowa, Iowa City, Iowa 52242, USA

I. Sample Details

All data presented in the main text are from point defects in hexagonal boron nitride (h-BN) flakes purchased from Graphene Supermarket. As received, the h-BN flakes are suspended in a 50/50 ethanol/water solution. Aside from the drop casting onto a thermally oxidized silicon substrate and the subsequent annealing procedure at $850°C$, the samples were not modified in any way. To verify that the investigated flakes are truly h-BN without traces of cubic boron nitride (cBN), for instance, we performed Raman spectroscopy. Figure S1a is an optical micrograph of the h-BN flakes after drop casting and annealing. The flake density here is 3x greater than for samples investigated in the main text. The flakes are clearly visible and they tend to aggregate. The boxed region in Figure S1a is magnified in Figure S1b. Spatially resolved Raman spectroscopy was performed on three locations denoted by the red, black, and blue circles. The Raman spectrum for each location is shown in Figure S1c. The red circle is in a region densely packed with flakes, and consequently its corresponding red Raman spectrum has a pronounced peak at the characteristic h-BN shift of ~$1366\ cm^{-1}$ (~170 meV). The black circle is in a location less dense with flakes and therefore a smaller Raman peak results. Lastly, because there are no flakes at the location of the blue dot, there is no h-BN Raman peak in the blue Raman spectrum. We did not observe any Raman peaks associated with cBN (~$1055\ cm^{-1}\ and\ 1305 cm^{-1}$) in our survey.

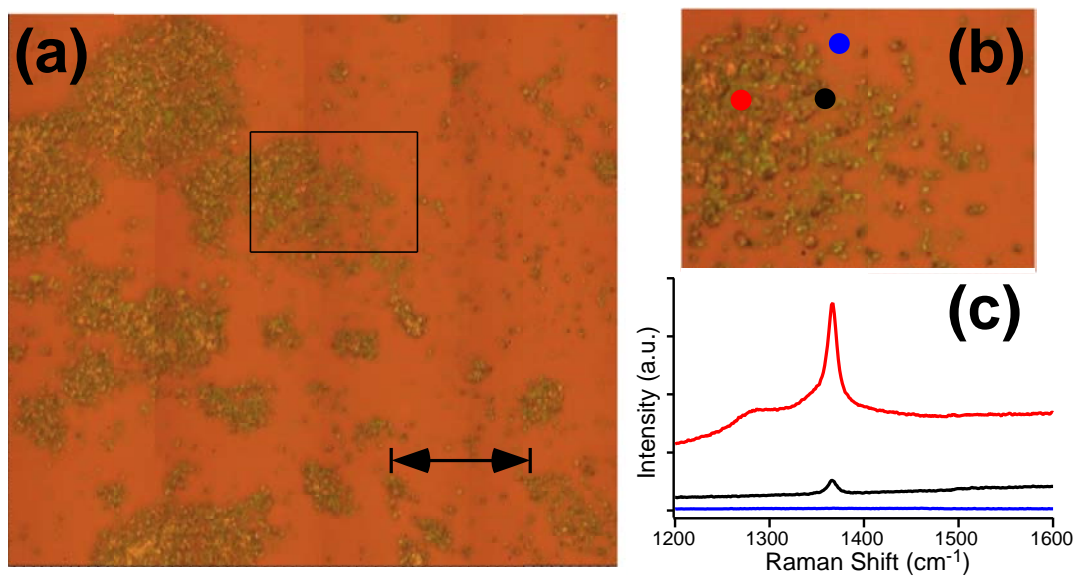

**Figure S1**: (a) Optical micrograph of h-BN flakes on a thermally oxidized Si substrate post annealing. The Scale bar is $30\ \mu m$. (b) Magnified view of the boxed region in (a). (c) Spatially resolved Raman spectra color-coordinated to the three colored circled in (b).

We also performed energy dispersive X-ray spectroscopy (EDX) of the h-BN flakes. Figure S2a is an electron micrograph of the flakes following drop casting and annealing. As in Figure S1a, the flakes are clearly visible as the brighter regions and they preferably aggregate. Figure S2b-e are elemental EDX maps of the boxed region for boron, nitrogen, oxygen, and carbon, respectively. As anticipated, the region corresponding to h-BN flakes shows high concentrations of boron and nitrogen whereas the region corresponding to the substrate is dominated by oxygen.

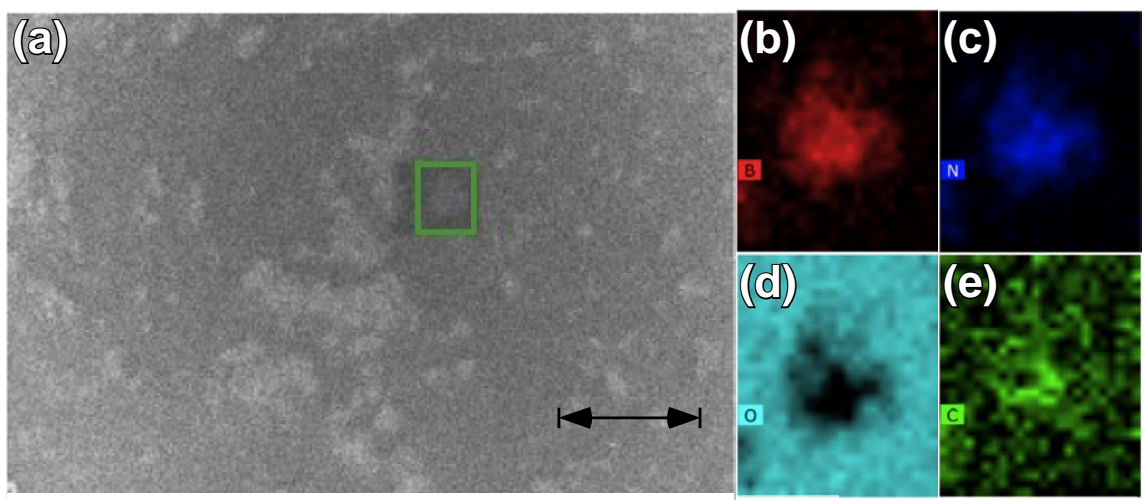

**Figure S2**: (a) Electron micrograph of h-BN flakes on a thermally oxidized Si substrate. The scale bar is $70\ \mu m$. Spatial EDX mapping for the boxed region in (a) for boron (b), nitrogen (c), oxygen (d), and carbon (e).

We also investigated defects in monolayer h-BN. These data are not in the main text but select results are presented in **Section III** below. These h-BN monolayers were grown on Cu foil via CVD and a 1 cm x 1 cm piece was subsequently transferred to a thermally oxidized Si substrate. For the transfer, PMMA was spin coated on the h-BN side of the as grown h-BN/Cu sample. The Cu foil was etched by dipping the sample in a Ferric Chloride solution (CE-100 from Transene Company Inc.) for 2-3 hours. Next, the PMMA/h-BN layer was placed in DI-Water and left to float for ~12 hours to remove all ionic residues that resulted from the Cu etching procedure. The PMMA/h-BN layer was then transferred to a thermally oxidized Si substrate and left to dry in air for 30-60 minutes. To remove the PMMA, the sample was submerged in acetone for ~12 hours. Finally, the h-BN monolayer was annealed in air at $500°C$ for 30 minutes.

## II. Experimental Apparatus

Figure S3 is a schematic of the house-built confocal microscope used in this work. A continuous wave 532 nm laser was used for the excitation source in all spectral and $g^{(2)}(\tau)$ measurements. For excitation measurements, a fixed polarizer (FP1) followed by a half wave plate (HWP1) enables creation of an arbitrary polarization state of the exciting light. A fixed wave plate (FWP1) corrects for retardances introduced by the rest of the excitation path.[1]

For the emitted light, a beam splitter (BS) placed before the dichroic mirror (DM) enables spectral measurements to be made that are minimally affected by wavelength-dependent transmission of optical elements. Photons not directed to the spectrometer encounter a Hanbury Brown and Twiss interferometer used for $g^{(2)}(\tau)$ measurements[2] and are either detected at APD1 or APD2. APD1 may be used for polarized excitation measurements and APD2 for measurements on the polarization of the emitted light. The filter wheel (FW) contains a combination of long- and short-pass filters that enable spectrally resolved time correlated single photon counting (TCSPC) measurements.

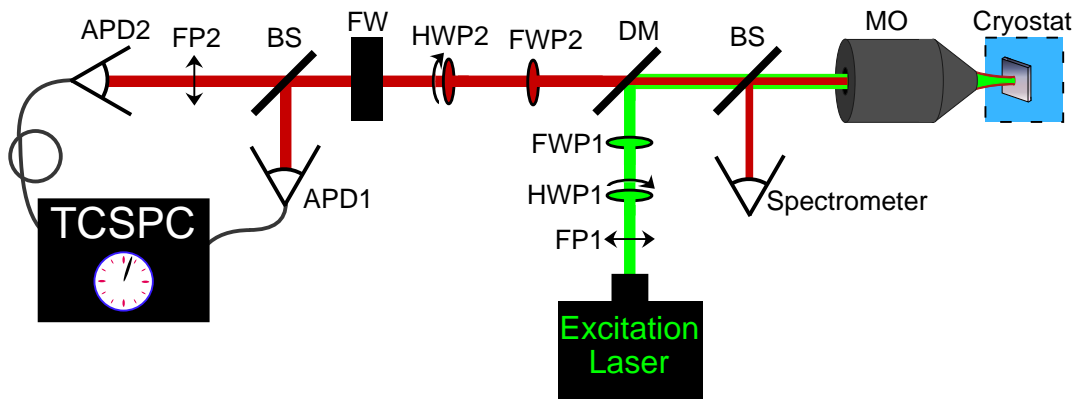

**Figure S3**: Schematic of confocal microscope used in this work.

For lifetime measurements the excitation source was a 532 nm pulsed laser (80 kHz repetition rate with 350 ps pulse width). Figure S4 displays two normalized lifetime measurements plotted on a log scale. The lifetime, $\tau$, of each transition is determined from an exponential fit, $\exp(-t/\tau)$, to the data.

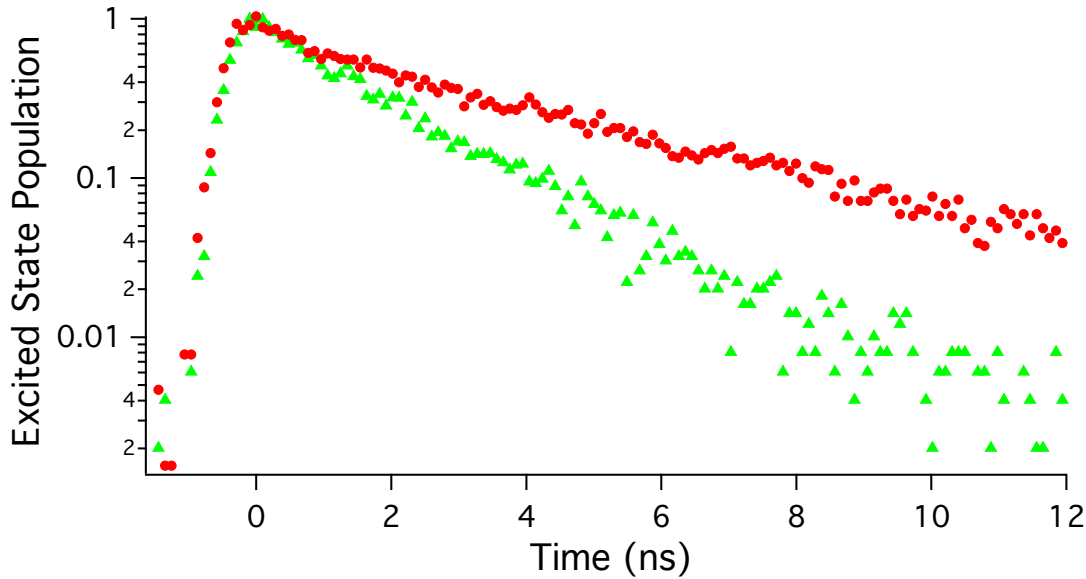

**Figure S4**: Normalized lifetime measurements acquired at 4.5 K for the 575 nm line (green triangles) and the 682 nm line (red circles).

### III. Point Defects in Monolayer h-BN

In contrast to the defects in multilayer h-BN presented in the main text, isolated defects in our single layer h-BN samples exhibit unreliable photostability and broad spectral emission. Figure S5 displays the fluorescence intensity of an isolated defect in single layer h-BN over ~14 seconds. Though bright, this defect blinks between a bright state and a dark state and eventually photobleaches after several minutes. Conversely, the multilayer h-BN defects discussed in the main text remained photoactive for the entire measurement duration (>24 hours for temperature dependent measurements) without bleaching.

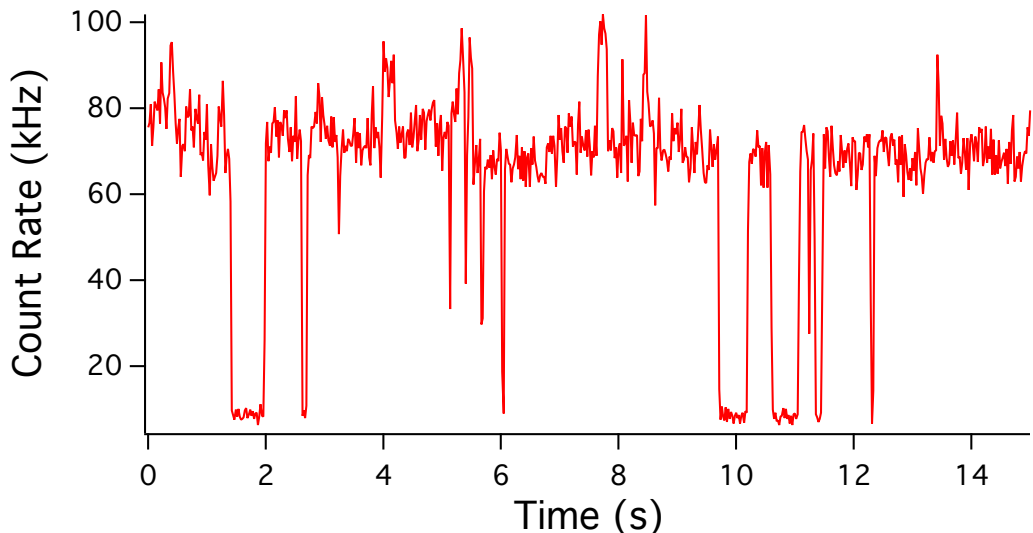

**Figure S5:** Fluorescence time trace of a single defect in monolayer h-BN.

Though defects in multilayer h-BN posses superior photostability, there is evidence that the defects may possess of the same structure. Figure S6 compares the emission spectra of a two distinct single defects in monolayer and multilayer h-BN. The monolayer emission is shown in red and was acquired at 80 K whereas the multilayer emission is shown in blue and was acquired at 150 K. Each spectrum has a maximum emission at ~575 nm, which facilitates direct comparison. The blue spectrum possesses a broadened zero-phonon line (ZPL) with an associated phonon sideband ~165 meV to the red. This energy is close to the Raman shift energy of ~170 meV (Figure S1) and also correlates well with the maxima of the h-BN phonon density of states.[3] The red spectrum from monolayer h-BN does not possess a narrow linewidth ZPL. Nonetheless, it likewise has a phonon sideband peak separated by the same energy of ~165 meV. These observations suggest that each defect may have the same structure.

If this is the case, then it may be possible to enhance the photostability of defects in monolayer h-BN via surface treatments.

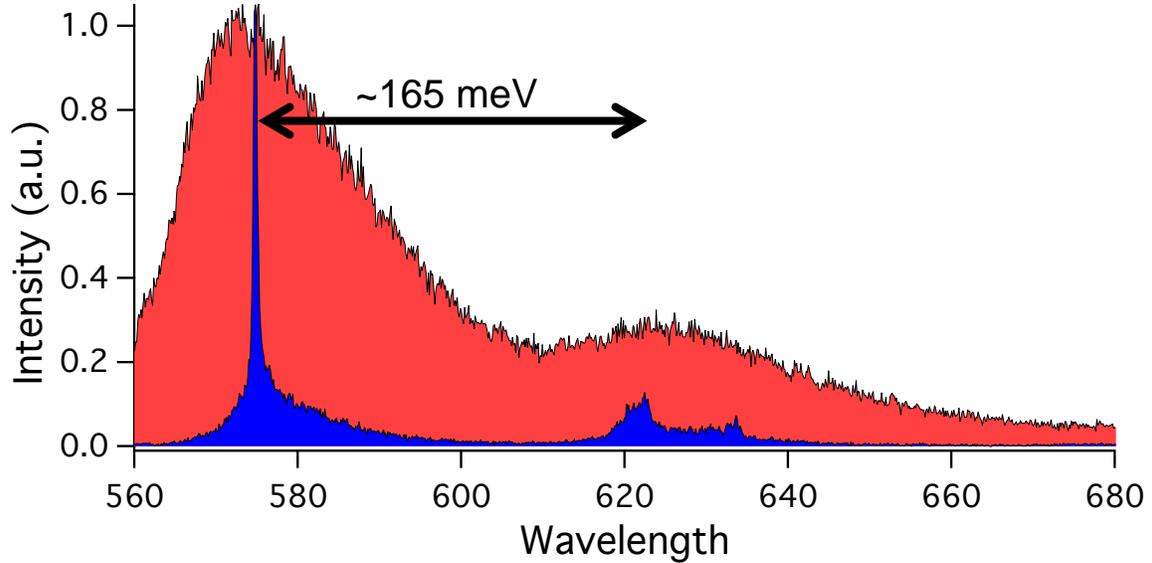

**Figure S6**: Comparison of the emission spectra of an individual defect in monolayer (red curve) and multilayer (blue curve) h-BN.

### IV. Emission Spectra Fitting and Relative Intensity

Each as-acquired emission spectrum, $I_\lambda(\lambda)$, represents the emission rate of photons with wavelength $\lambda$ per unit wavelength. In this work, each as-acquired emission spectrum was first converted to energy spectral density, $I_E(E)$, according to the relation $I_E(E) \propto \lambda^2 I_\lambda(\lambda)$. As stated in the main text, sharp emission lines were then fit to a linear combination of a background and a Lorentzian lineshape. The FWHM of each line was extracted directly from the FWHM of the Lorentzian fit. We define the relative intensity of each line, $I_R$, as the ratio of the Lorentzian lineshape intensity to that of the background at $E_{zpl}$. Figure S7 illustrates the how the relative intensity is calculated. Here the blue curve is the 575 nm ZPL at 360 K and the red curve is the same data with the

Lorentzian fit subtracted. The dashed line corresponds to $E_{zpl}$ and the blue and red diamonds correspond to $I_{tot}(E_{zpl})$ and $I_{BG}(E_{zpl})$, where $I_{tot}$ is the total signal and $I_{BG}$ is the background signal. Note that the background signal may arise from other defects, the substrate, or from multiphonon processes of the same defect that produced the 575 nm line. In Figure S7 the relative intensity becomes $I_R = [I_{tot}(E_{zpl}) - I_{BG}(E_{zpl})]/I_{BG}(E_{zpl}) \approx 1.5$.

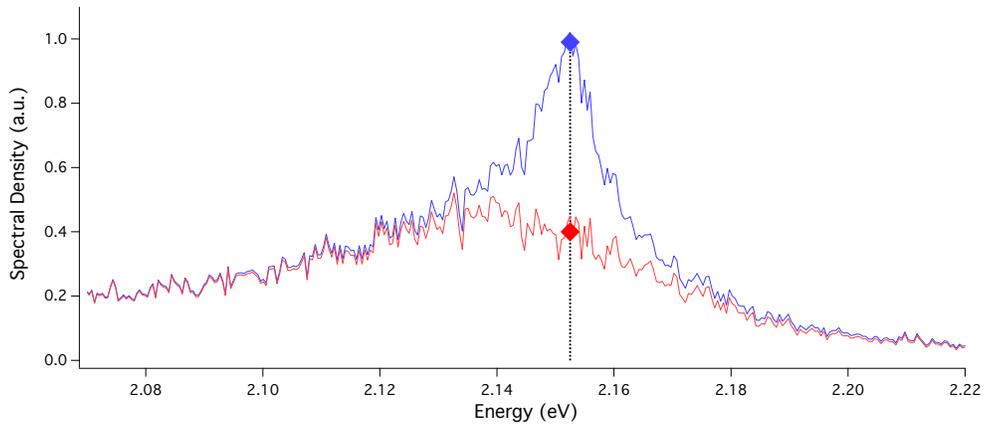

**Figure S7**: Energy spectral density of 575 nm line at 360 K before (blue) and after (red) subtraction of the Lorentzian fit.

## V. Linewidth Broadening Model and Fitting

We assume that near the zero-phonon line the spectral lineshape, $G(E)$, solely results from zero-phonon and one-phonon processes:

$$G(E) = G_{zpl}(E) + G_{opl}(E). \qquad (1)$$

Here $G_{zpl}$ is the true zero-phonon lineshape, $G_{opl}$ is the lineshape that results from one-phonon processes, and $E$ is the energy shift from the zero phonon line energy, $E_{zpl}$. In the Gaussian limit, the full width at half maximum (FWHM), $\Gamma$, of $G(E)$ is determined from the second moment:

$$\Gamma^2 = 8\ln(2)\frac{\int E^2 G(E)dE}{\int G(E)dE} \tag{2}$$

Experimentally, the measured linewidth, $\Gamma_M$, is limited by the device resolution $R$:

$$\Gamma_M = \sqrt{8\ln(2)\frac{\int E^2 G(E)dE}{\int G(E)dE} + R^2} \tag{3}$$

In the weak coupling limit,[4] $G_{opl}$ is given by:

$$G_{opl}(E) = \frac{cg(E)f(E)}{\exp\left(\frac{E}{kT}\right)-1}. \tag{4}$$

Here $g(E)$ is the phonon density of states, $f(E)$ describes the defect-phonon coupling, $k$ is Boltzmann's constant and $T$ is the temperature. Additionally, the spectral weight of the ZPL diminishes with temperature as:

$$\int G_{zpl}(E)dE = D\exp\left(-S\mathrm{Coth}(\hbar\omega_0/2kT)\right). \tag{5}$$

Here $S$ is the Huang-Rhys factor, $\hbar$ is the reduced Planck's constant, $\omega_0$ is the phonon frequency, and $D$ is a constant. By combining Equations 3-5, and neglecting the natural width of the ZPL, we obtain:

$$\Gamma_M = \sqrt{8\ln(2)\frac{\int E^2 G_{opl}dE}{D\exp[-S\mathrm{Coth}(\hbar\omega_0/2kT)]+\int G_{opl}(E)dE} + R^2}. \tag{6}$$

In the limit that low-energy phonons dominate, we have

$$\Delta E = E_{exc} - E_{zpl} = 2S\hbar\omega_0, \tag{7}$$

where $E_{exc}$ is the laser excitation energy. Then Equation 6 becomes:

$$\Gamma_M = \sqrt{8\ln(2)\frac{\int E^2 G_{opl}dE}{D\exp[-S\mathrm{Coth}(\Delta E/4SkT)]+\int G_{opl}(E)dE} + R^2}. \tag{8}$$

Equation 4 and 8 jointly model the effective thermal broadening of the emission line by low-energy acoustic phonons. What remains to be determined are the bounds of integration, the explicit form of the phonon density of states, $g(E)$, and the defect-phonon coupling term $f(E)$.

In two dimensions with piezoelectric coupling, $g(E) \propto E$ and $f(E) \propto E^{-1}$. In such a scenario our fitting function of two free parameters $A, S$ becomes:

$$\Gamma_M = \sqrt{8\ln(2) \frac{\int E^2 dE / \left[\exp\left(\frac{E}{kT}\right) - 1\right]}{A\exp[-S\coth(\Delta E/4SkT)] + \int dE / \left[\exp\left(\frac{E}{kT}\right) - 1\right]} + R^2}. \tag{9}$$

We used Equation 9 to fit our experimentally measured linewidths using a $\chi^2$ minimization procedure. In our best-fits, the integration bounds in Equation 9 were 0.001 meV to 200 meV. The 200 meV cutoff was chosen because this is where the h-BN phonon density of states vanishes.[3] The spectrometer resolution $R$ was determined to range between 0.3 meV and 0.4 meV depending on the spectrometer slit width.

To test our assumption that in-plane phonons with piezoelectric coupling dominate the measured linewidth we also considered the following fitting function:

$$\Gamma_M = \sqrt{8\ln(2) \frac{\int E^{2+n} dE / \left[\exp\left(\frac{E}{kT}\right) - 1\right]}{A\exp[-S\coth(\Delta E/4SkT)] + \int E^n dE / \left[\exp\left(\frac{E}{kT}\right) - 1\right]} + R^2}. \tag{10}$$

When fitting using Equation (10), we fixed $n$ and allowed $A, S$ to vary as free parameters. In Figure S8 we plot the normalized minimum $\chi^2$ for different values of $n$ for each ZPL investigated. This plot supports the value $n = 0$ and our assignment that the relevant phonons are in-plane with piezoelectric coupling.

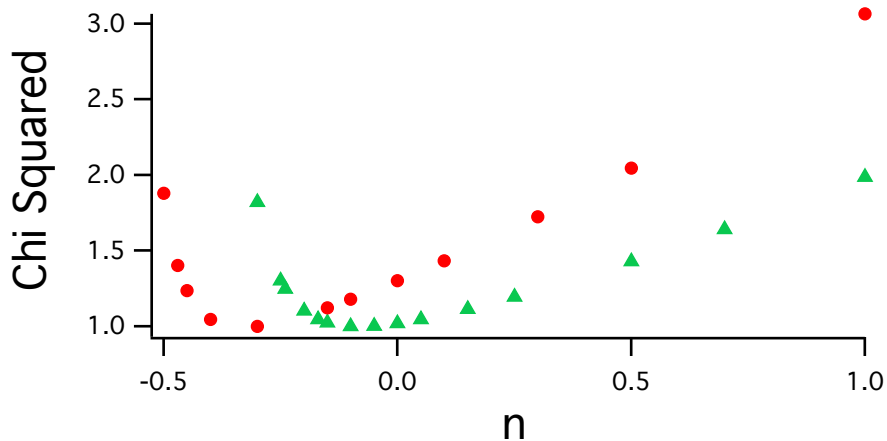

**Figure S8**: Minimum $\chi^2$ for different values of $n$ for the 575 nm ZPL (green triangles) and the 682 nm ZPL (red circles).

**References:**

(1) Jungwirth, N. R.; Chang, H.-S.; Jiang, M.; Fuchs, G. D. Polarization Spectroscopy of Defect-Based Single Photon Sources in ZnO. *ACS Nano* **2016**, *10*, 1210–1215.
(2) Jungwirth, N. R.; Pai, Y. Y.; Chang, H. S.; MacQuarrie, E. R.; Nguyen, K. X.; Fuchs, G. D. A Single-Molecule Approach to ZnO Defect Studies: Single Photons and Single Defects. *J. Appl. Phys.* **2014**, *116*, 043509.
(3) Tohei, T.; Kuwabara, A.; Oba, F.; Tanaka, I. Debye Temperature and Stiffness of Carbon and Boron Nitride Polymorphs from First Principles Calculations. *Phys. Rev. B* **2006**, *73*.
(4) Stoneham, A. M. *Theory of Defects in Solids: Electronic Structure of Defects in Insulators and Semiconductors*; OUP: Oxford, 2001.